\documentstyle[epsfig]{kapproc} 
\kluwerbib
\startauthorindex

\def\lsim{\vcenter{\hbox{$<$}\offinterlineskip\hbox{$\sim$}}}

\begin{document}

\articletitle{Young stars and their circumstellar disks in the $\sigma$\,Orionis cluster}


\vspace*{-0.4cm}
\author{J.M. Oliveira, R.D. Jeffries, J.Th. van Loon, M.J. Kenyon}
\affil{School of Chemistry and Physics, Keele University, Staffordshire ST5 5BG, UK}

\chaptitlerunninghead{Young stars and their circumstellar disks in $\sigma$\,Orionis cluster}

\anxx{Oliveira et al.}
\vspace*{-0.4cm}
\begin{abstract}
The $\sigma$\,Orionis cluster is a young association evolving under the 
disruptive influence of its massive O-star namesake. We are analysing 
this cluster as part of a program to characterise the influence of O-stars on the early stages of stellar evolution. At an age of approximately 4\,Myr, this cluster is at a crucial stage in terms of disk evolution and therefore it is a key case to better constrain disk dissipation timescales. We have obtained $RI$ photometry and optical spectroscopy of the $\sigma$\,Ori cluster; we have analysed the Li\,{\sc i} and Na\,{\sc i} features to establish cluster membership. We have thus gathered a unique sample of spectroscopically confirmed low-mass cluster members and brown dwarfs. Disk frequencies from K-band excesses from 2MASS suggest that less than 7\% of the very low-mass $\sigma$\,Ori members have disks (Oliveira et al. 2002), in stark contrast with even younger clusters (e.g. Trapezium). However, near-infrared disk frequencies have to be taken with caution. We are currently undertaking an L-band (imaging) and mid-infrared (imaging and spectroscopy) program to identify and probe the properties of circumstellar disks around young stars in this cluster. Preliminary results indicate that at least 30\% of cluster members have circumstellar disks.

\end{abstract}

\section*{Introduction}

\paragraph{Low-mass stars in OB associations} Historically, OB associations were identified as loose groups of coeval O-and B-type stars. However when extrapolating the mass function for OB stars to lower masses, it was found that most of the mass in such groups should actually be in low-mass stars ($\lsim 2$\,M$_{\odot}$) (e.g.\ Brown 2001). Indeed, H$\alpha$ objective prism surveys and X-ray surveys unveiled large populations of low-mass stars in these associations. \\
\indent Most stars may form in such associations (Walter et al. 2001), with massive and low-mass stars originating in the same star forming regions and at approximately the same epoch. If this is the case, then the presence of these luminous stars in the midst of lower mass siblings raises important questions: do OB stars influence the rate of lower-mass star formation and do they affect pre-main-sequence (PMS) stellar evolution and, more crucially, circumstellar disk evolution?
 
\paragraph{The importance of circumstellar disks} Circumstellar disks are an important part of the star formation process and are ubiquitous around young stars. Even though they seem to be relatively short lived, they impact strongly on stellar evolution. The interaction between the disk and the stellar magnetic field plays a central role in angular momentum regulation and it is thereby likely to influence the spread of rotation rates in young stars. Young stars may also accrete a significant fraction of their final mass from their disks, so disk evolution could impact strongly on theoretical PMS models and attempts to determine stellar masses and ages.\\
\indent Two very important questions are disk dissipation timescales and its possible mass dependence. Disk dissipation timescales might be the strongest factor in determining the timescales for planet formation or whether planets form at all in a particular stellar system. This could be even more crucial in OB associations where there are exterior sources of photo-evaporating radiation. Mass dependence of disk dissipation could be important in the context of brown dwarf formation. It has been proposed that brown dwarfs form by ejection from their parent systems; numerical simulations (Bate et al. 2003) propose that the dynamical interactions that eject brown dwarfs also truncate their disks, making them relatively small and of low mass. A testable prediction of such models is that brown dwarf disks dissipate faster. 

\paragraph{The $\sigma$\,Orionis cluster} $\sigma$\,Orionis is a Trapezium-like system with an O9.5\,V primary. The population of low-mass stars spatially clustered around this system was discovered as bright sources in ROSAT X-ray images, and follow-up optical spectroscopy confirmed most sources as PMS stars (Walter et al. 1997). This association is young, nearby and affected by low reddening, making it an ideal target to analyse the PMS population, even down to brown dwarfs and isolated planetary mass objects. The main goal of the work we summarise here is to probe the properties of the PMS population in the $\sigma$\,Ori cluster, establishing a representative sample of cluster members and analysing the frequency and properties of circumstellar disks.

\section*{Low-mass pre-main-sequence population}

\begin{figure}[h]
\hspace*{-0.5cm} \centerline{\psfig{file=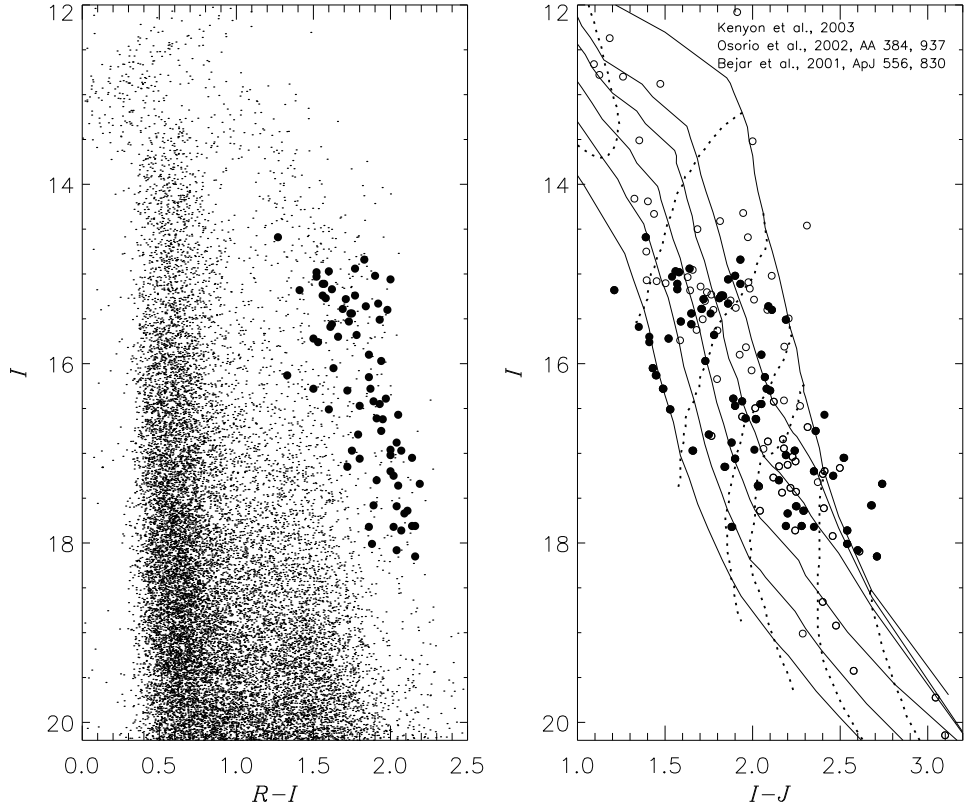,height=11.5cm}}
\caption{Left: $RI$ colour-magnitude diagram around $\sigma$\,Orionis. Right: $IJ$ colour-magnitude diagram for the cluster members, with model tracks (for 0.8, 0.5, 0.2, 0.1, 0.075, 0.05 and 0.03\,M$_{\odot}$) and isochrones (for 1, 3, 5, 10, 20 and 50\,Myr) from Baraffe et al. (1998). The objects identified by us (Kenyon et al. 2003 in preparation) are represented by filled symbols; empty symbols are objects identified by other authors. }
\end{figure}

\paragraph{Photometric surveys} Firstly, it is necessary to identify the low-mass PMS content near $\sigma$\,Ori. Fig.\,1 (left) shows an $RI$ colour-magnitude diagram, obtained with the Wide Field Camera (WFC) at the Isaac Newton Telescope (INT), where the PMS stars can be seen separated from the bulk of the contamination. It was based on this diagram that targets were identified for follow-up spectroscopy. The figure also shows (right) the $IJ$ colour-magnitude diagram for cluster members --- J-magnitude comes from the 2MASS database. It shows most of the (over 100) cluster members identified so far, either by us (see below) or from the literature (e.g.\ B\'{e}jar et al. 2001; Zapatero Osorio et al. 2002). Overplotted are the model tracks and isochrones from Baraffe et al. (1998). We estimate that the median age of the cluster is approximately 4\,Myr (Oliveira et al. 2002) and the sample covers the mass range 0.02$-$1\,M$_{\odot}$. It is worth pointing out that a large number of brown dwarfs have been identified (30$-$40 depending on the adopted age). Several isolated planetary mass objects ($M \leq 0.013$\,M$_{\odot}$) have been discovered in the cluster (e.g.\, Zapatero Osorio et al. 2000).

\paragraph{Membership Analysis} One of the main drives of our spectroscopic observations was to investigate how efficient photometric surveys are in identifying PMS stars and thus cluster members. This is actually quite important for the disk frequency analysis as contamination leads to lower disk frequencies. \\
\indent We have performed spectroscopic observations of approximately 70 photometric cluster candidates with WYFFOS at the William Hershel Telescope (WHT). We use 3 means of diagnostic for cluster membership: radial velocity measurements (compared with the cluster mean of $\sim$\,32\,km\,s$^{-1}$), the equivalent width of the Li\,{\sc i} 6708\,\AA\ feature (a fragile element that remains unburned for stars younger than 10\,Myr), and the equivalent width of the Na\,{\sc i} 8190\,\AA\ doublet (a gravity indicator that is very weak in the spectra of giant stars, is strong for field dwarfs and weak for PMS stars). Using these indicators together we are able to firmly identify cluster members and binary candidates. We found that 90\% of the objects in the photometric sample are indeed cluster members. This implies that photometric surveys are quite efficient in identifying cluster members. This analysis is described in Kenyon et al. (2003 in preparation).

\section*{Circumstellar disks in the $\sigma$\,Orionis cluster}

We have an on-going program to survey the circumstellar disk population in the $\sigma$\, Ori cluster. The traditional first step is to look for K-band excesses in our sample of cluster members. Using $JHK$ photometry from 2MASS, it has been established (Oliveira et al. 2002; Barrado y Navascu\'{e}s 2003) that at most 10\% of the cluster members exhibit a K-band excess that could indicate the presence of a circumstellar disk. However, a K-band excess is an unreliable disk indicator. It depends strongly on the inner disk temperature; in particular for the lower mass stars, there is evidence that their disks are not hot enough to produce significant K-band excess (Natta \& Testi 2001; Comer\'{o}n et al. 2000).

\subsection*{L-band survey with UIST/UKIRT}

\begin{figure}[ht]
\centerline{\psfig{file=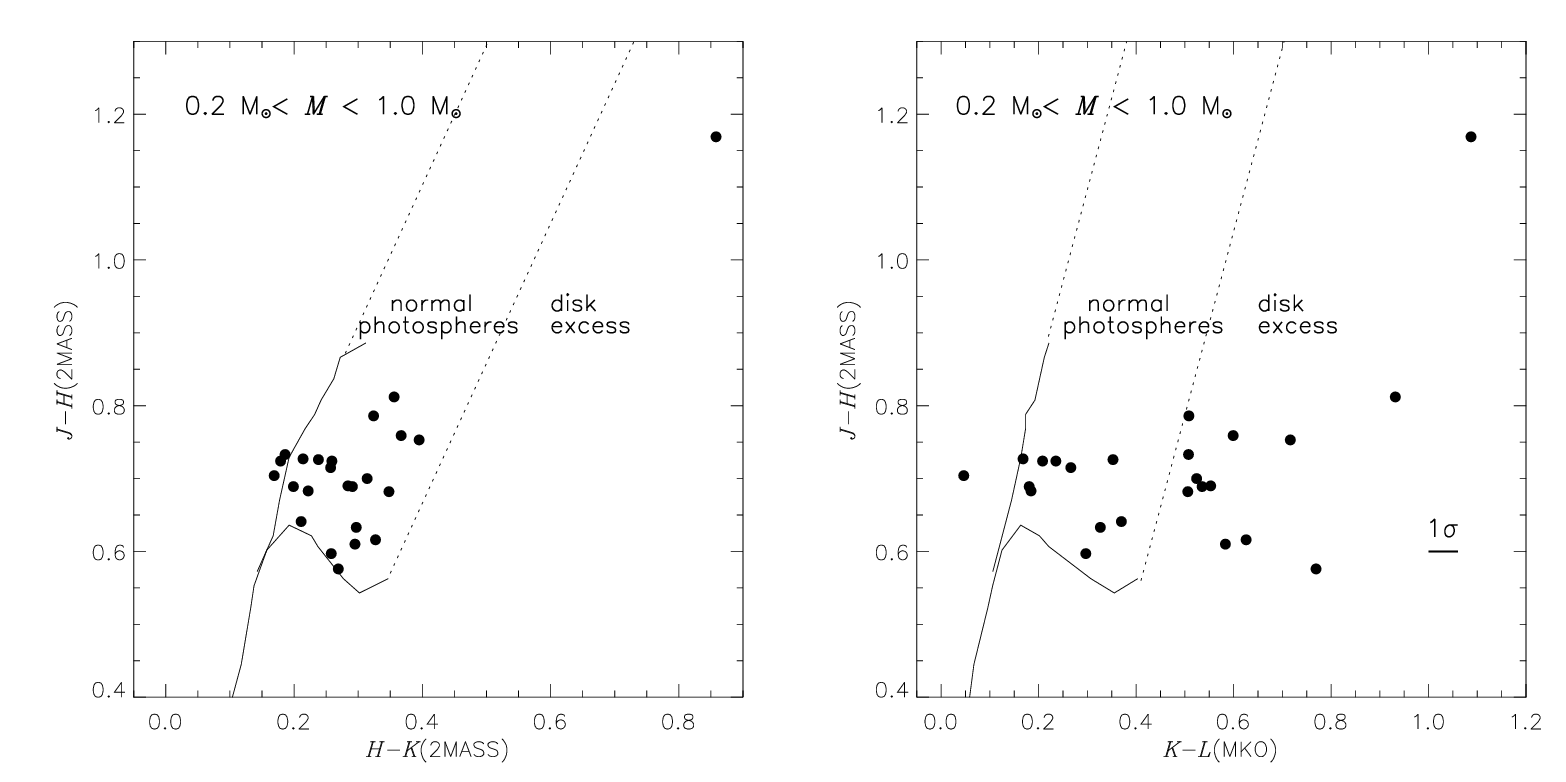,height=7.8cm}}
\centerline{\psfig{file=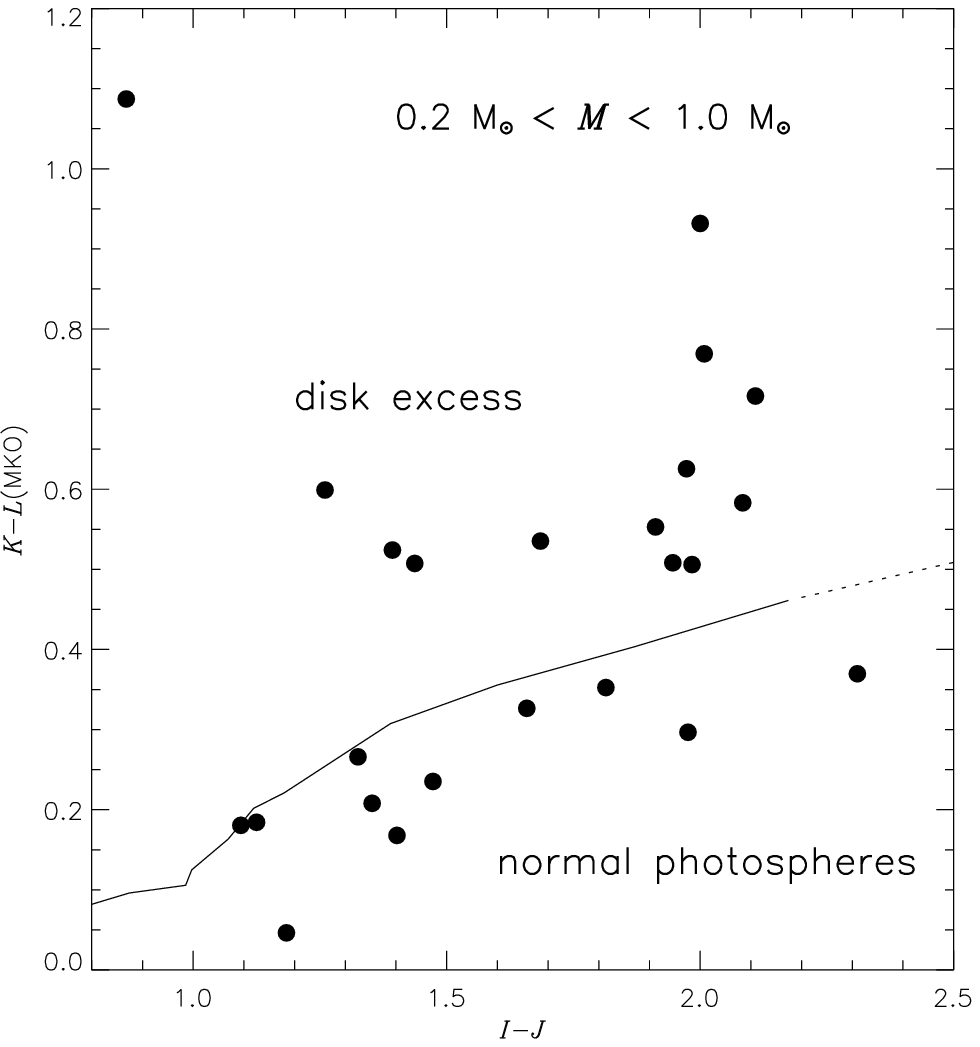,height=7.8cm}}
\caption{$JHK$, $JHKL$ and $IJKL$ colour-colour diagrams for the $\sigma$\,Ori sample observed at UKIRT. As expected, most objects do not show an $H-K$ excess (top left) but at least 30\% exhibit a $K-L$ excess (top right) that would indicate a circumstellar disk. This is re-enforced by the $IJKL$ diagram (bottom) that is almost insensitive to variability (in the I- and J-bands) and reddening.}
\end{figure}

At longer wavelengths, infrared excesses grow rapidly thus L-band observations are the most reliable and efficient way to detect circumstellar disks, down to the substellar regime. We have a program to observe all known $\sigma$\,Ori cluster members in the L-band (and also in the K-band to avoid uncertainties related with variability, Carpenter et al. 2002); we have obtained data in January 2003 with the newly installed UIST at the UK Infrared Telescope (UKIRT). Due to adverse weather conditions we could only perform about 30\% of the program, observing the targets in the mass range 0.2$-$1.0\,M$_{\odot}$. We have since then been allocated more observing time at UKIRT to complete this program.\\
\indent Fig.\,2 shows colour-colour diagrams for the sample observed at UKIRT. On the top left is the 2MASS $JHK$ diagram; only one object has a significant K-band excess, indicative of a circumstellar disk. The situation changes dramatically in the $JHKL$ diagram (top right) where many stars seem to exhibit $K-L$ excess, evidence of circumstellar disks. A very revealing diagram is $IJKL$ (bottom). Both photometric variability in I- and J-band and reddening would move an object's position almost horizontally in this diagram, while a disk excess moves an object vertically. This means that the diagnostic value of this diagram for the detection of disks is for all effects insensitive to variability and reddening. At least 30\% of the objects seem to have an excess that indicates the presence of a circumstellar disk, but this value might be as high as 50\%. Only the computation of the intrinsic colour for each object will reveal the true percentage of objects with an excess (Oliveira et al. 2003 in preparation).

\subsection*{Mid-infrared survey with TIMMI2/ESO}

\begin{figure}[ht]
\centerline{\psfig{file=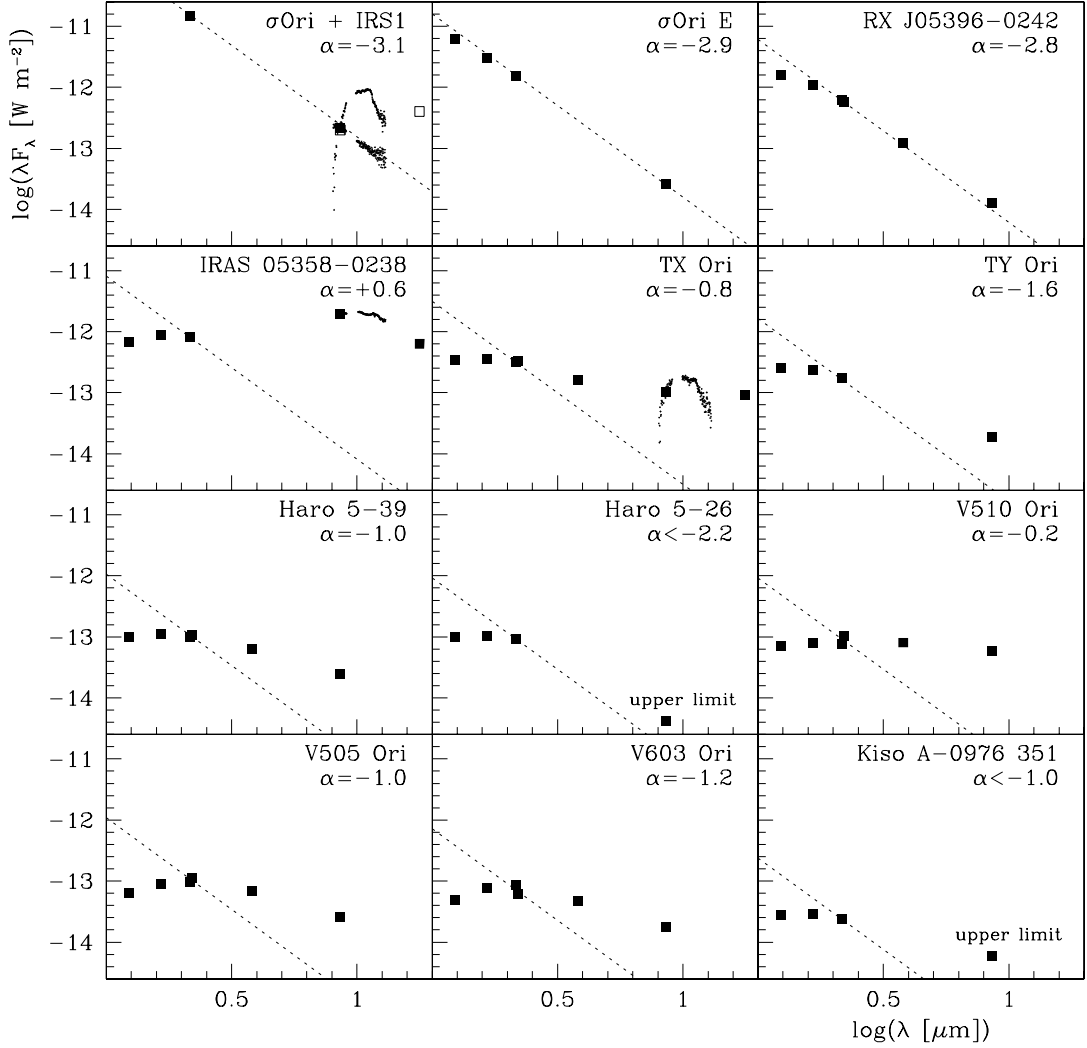,height=15.5cm}}
\caption{SEDs for the TIMMI2 targets. The identification of each source is given on the top right corner of each graph as well as the spectral index $\alpha$. The objects on the top row have earlier spectral types and their SEDs show no excess emission at mid-infrared wavelengths. All the other objects show excesses consistent with circumstellar disks. For some objects we also have spectra around the 10\,$\mu$m silicate emission feature.}
\end{figure}

L-band observations are adequate for disk surveys, but in order to characterise circumstellar disk properties, observations at even longer wavelengths are necessary. Based on estimates from the original IRAS data for the possible mid-IR emission of PMS stars around $\sigma$\,Ori, we have targetted 12 such objects with TIMMI2 at the ESO 3.6\,m telescope to obtain N-band photometry, as well as Q-band photometry and N-band spectroscopy for the brightest amongst them.\\
\indent Fig.\,3 shows the spectral energy distribution (SED) of these sources. Indicated in the upper right corner of each graph is the spectral index $\alpha$\ that is a measure of the mid-IR excess. The three objects on the top row, $\sigma$\,Ori itself, $\sigma$\,Ori E and a ROSAT source are the more massive, early spectral type objects. Their SEDs are consistent with the slope of the Rayleigh-Jeans tail of a blackbody ($\alpha \sim 3$), so there is no excess emission. The other objects all show an excess at these wavelengths, evidence of circumstellar disks. For instance, TX Ori has $\alpha =-0.8$ consistent with a Class\,II classification (classical T Tauri star, CTTS); the spectrum in the N-band reveals a feature consistent with silicate dust emission at 10\,$\mu$m. Based on the value of $\alpha$, most objects can be classified as Class\,II objects. Two objects exhibit SEDs that suggest more complex circumstellar structures: V510\,Ori (associated with a jet) has $\alpha$ consistent with a flat-spectrum source and IRAS\,05358-0238 has $\alpha$ consistent with a Class\,I source (an intermediate state between an embedded Class\,0 source and a CTTS). This analysis is described in full in Oliveira \& van Loon (2003 in preparation).

\section*{Final remarks}

We have combined photometric and spectroscopic surveys and identified a very significant number of cluster members, that together with other surveys constitutes a unique sample of PMS objects, covering the mass range 0.02$-$1.0\,M$_{\odot}$. \\
\indent We have begun a program to search for cluster members with circumstellar disks in the L-band. Preliminary results on a representative sample (0.2$-$1.0\,M$_{\odot}$) indicate that at least 30\% of $\sigma$\,Ori cluster members have circumstellar disks. When compared with two other clusters of similar ages (NGC\,2264 at $\sim$\,3.2\,Myr and NGC\,2362 at $\sim$\,5\,Myr), a disk frequency of 30\% for the $\sigma$\,Ori cluster would fit nicely between these two clusters (disk frequencies respectively 52\% and 12\%), broadly agreeing with the 6\,Myr timescale for all stars to lose their disk as proposed by Haisch et al. (2001). As our L-band survey does not yet reach low enough masses, we are not able to address the issue of mass dependence of disk frequency. We intend to continue our program to search for and characterise circumstellar disks around fainter members of the $\sigma$\,Ori cluster.


\end{document}